%% file: main.tex
\documentclass[sigconf, authorversion=true]{acmart}

\usepackage{booktabs} 

\usepackage[caption=false,font=footnotesize]{subfig}
\usepackage{fancyhdr}
\usepackage{mdwlist}
\usepackage{multirow}
\usepackage{amsmath}        

\graphicspath{{incl/}}

\makeatletter
\def\blfootnote{\xdef\@thefnmark{}\@footnotetext}
\makeatother

\usepackage{setspace}
\def\smallerspacecaption{\vspace{-2mm}}

\definecolor{gray}{gray}{0.9}

\newcommand{\drop}[1]{}

\begin{document}

\title{Raise Your Game for Split Manufacturing:\\
		Restoring the True Functionality Through BEOL
}

\author{Satwik Patnaik, Mohammed Ashraf, Johann Knechtel, and Ozgur Sinanoglu}
 \affiliation{%
   \institution{Tandon School of Engineering, New York University, New York, USA}
   \institution{Division of Engineering, New York University Abu Dhabi, United Arab Emirates}
 }
 \email{{sp4012, ma199, johann, ozgursin} @nyu.edu}

\renewcommand{\headrulewidth}{0.0pt}
\thispagestyle{fancy}
\pagestyle{fancy}

\begin{abstract}
\input{abstract}
\end{abstract}

\copyrightyear{2018} 
\acmYear{2018} 
\setcopyright{acmcopyright}
\acmConference[DAC '18]{DAC '18: The 55th Annual Design Automation Conference 2018}{June 24--29, 2018}{San Francisco, CA, USA}
\acmBooktitle{DAC '18: DAC '18: The 55th Annual Design Automation Conference 2018, June 24--29, 2018, San Francisco, CA, USA}
\acmPrice{15.00}
\acmDOI{10.1145/3195970.3196100}
\acmISBN{978-1-4503-5700-5/18/06}

\maketitle

\renewcommand{\shortauthors}{S.\ Patnaik et al.}

\section{Introduction}
\label{sec:introduction}

Proposed in 2011 by the IARPA agency,
\emph{split manufacturing (SM)}
seeks to protect chip design companies against piracy of their intellectual property (IP) by third-party manufacturing facilities~\cite{mccants11}. SM can also help to mitigate related threats such as insertion of hardware Trojans or unauthorized
over-manufacturing~\cite{HW_Obfuscation-FBT17}.

In the most common threat model~\cite{rajendran13_split,hill13,wang16_sm,magana16,magana17,
sengupta17_SM_ICCAD,feng17},
the front-end-of-line (FEOL)
is handled by an outsourced, high-end fab which is considered competitive but \emph{untrustworthy}, whereas the back-end-of-line (BEOL) is subsequently manufactured (on top of the
		FEOL) at a
\emph{trusted} integration facility.
Besides, Wang \emph{et al.}~\cite{wang17_FEOL} addressed
another threat model where the BEOL facility is untrusted; here the adversaries seek to infer the gates
from the whole BEOL stack (i.e., all wires/vias are available, except the intra-cell wiring in M1).

The security promise of SM is based on two assumptions: ($i$)~third-party manufacturers do not have access to the complete design but only to
either FEOL or BEOL and
($ii$)~those third parties are not colluding.
For the
conventional threat model (untrusted FEOL fab), there is an
additional risk: adversaries in the fab can leverage the physical implementation details of the FEOL layout. 
More specifically, since design automation tools optimize
for power, performance, and area (PPA), the FEOL part itself
	contains various hints on the missing
	BEOL interconnects. Most importantly, gates to be connected are typically placed close to
each other. 
This hint on proximity (among others such as delay constraints or routing paths) is leveraged by 
various \emph{proximity attacks}~\cite{rajendran13_split,wang16_sm,magana16,magana17,feng17}, raising concerns regarding the security
promise offered by SM.

In this work, we ``raise the game'' for SM to protect against malicious fab adversaries. The fundamental idea is to manipulate both placement and routing in an
intertwined, holistic and misleading manner, thereby increasing the resilience of FEOL layouts.
More specifically, we randomize the design at the netlist level,
		place and route the resulting erroneous netlist, and restore the true functionality only in the BEOL. This paper
can be summarized as follows:

\begin{itemize}

\item We initially review state-of-the-art proximity attacks, related metrics, prior protection schemes, and associated short-comings (Sec.~\ref{sec:background}). We also outline how
our concept can
mislead proximity attacks, to begin with, even for attacks
that presumably achieve perfect scores (Sec.~\ref{sec:concept}).

\item We propose and implement (in \emph{Cadence Innovus}) a protection scheme for SM (Sec.~\ref{sec:method}). 
Our scheme is based on holistic placement and routing perturbation in the FEOL and subsequent
correction in the BEOL. The scheme allows for controllable
impact on PPA.
       Moreover, we can limit the commercial cost of SM, 
       by splitting after higher layers (e.g., after M6).

\item 
We design custom \emph{correction cells} for our scheme.
These cells allow us to handle wire detours in a well-controlled manner, which is essential to ($i$) induce misleading placement
and routing in the FEOL and ($ii$) restore the true functionality later on by re-routing in the BEOL.

\item We evaluate our scheme in terms of PPA cost and security (Sec.~\ref{sec:results}).
We consider various benchmarks, including the industrial \emph{IBM superblue} benchmarks. 
We thoroughly contrast
with prior art. We also make our
DRC-clean protected layouts and SM scripts available to the community, along with the
library definitions for the correction cells~\cite{webinterface}.

\end{itemize}

\section{Background and Motivation}
\label{sec:background}

Wang \emph{et al.}~\cite{wang16_sm} proposed an advanced proximity attack which utilizes multiple hints from the FEOL layouts: ($i$) physical proximity of gates, ($ii$) avoidance
of combinatorial loops, ($iii$) constraints on
load capacitances, ($iv$) direction of ``dangling wires''\footnote{There are metal segments left open/unconnected in the topmost FEOL layer,
	namely where the vias connecting upward
to the BEOL are to be placed. These metal segments are referred to as ``dangling wires.''
Moreover, we refer to the locations for those vias connecting with the BEOL as \emph{virtual pins (vpins)}, as in~\cite{magana16,magana17}.}, and ($v$) timing constraints.
Maga\~{n}a \emph{et al.}~\cite{magana16,magana17}
proposed different attack schemes, whereupon they empirically observe that attacks considering routing paths/utilization are more effective than placement-centric attacks.
They also observe that
the \emph{IBM superblue} suite is considerably more challenging to attack than ``traditional,'' small-scale benchmarks.
Note that their attacks
do not recover actual
netlists, but only list 
	possible candidates for each net to reconnect.

Key attack metrics, as discussed in~\cite{rajendran13_split,wang16_sm, wang17}, are the \emph{Hamming distance (HD)}, the \emph{correct connection rate (CCR)}, and the \emph{output error rate (OER)}. 
The HD quantifies the mismatch between the outputs of an original and the outputs of a recovered/stolen netlist during test stimulation. 
An HD of 0\% (or 100\%) 
denotes attack success.
The CCR is the ratio of successfully recovered nets over all protected nets. 
Hence, the higher the CCR, the more effective the attack.
The OER reflects on the probability of some output bits being incorrect when stimulating the recovered/stolen netlist with test patterns.
The routing-centric metrics proposed in~\cite{magana16,magana17}
gauge the solution space of SM.
The \emph{number of vpins} counts the overall vias/pins in the topmost FEOL layer (which are to be reconnected);
the \emph{candidate list size} is the average number of nets to consider for each vpin;
and the \emph{match in list} reflects for how many vpins the correct net is among those in the candidate list.\footnote{
Consider the following example, where an attacker
observes 1,000 vpins. In accordance with~\cite{magana16}, also consider that only two-pin nets are
available in the design, i.e., 500 two-pin nets are to be reconnected.
The size of the solution space covering all possible netlists 
is the number of perfect matchings in a complete bipartite graph
(representing the 500 drivers and 500 sinks), which is simply
$500! = 1.22\times10^{1143}$.
After conducting the routing-centric attacks,
      assuming the best possible \emph{match in list} (i.e., 100\%) and a \emph{candidate list size} of 1.4 on average, there are
at most ``only'' $\approx 1.4^{500} = 1.16\times10^{73}$ possible netlists remaining. (This number is coincidentally approaching the estimated number of atoms in the universe.)
It is important to note that for any match in list below 100\%,
the true netlist is not even covered by that large number.}
Thus, while the related attacks
help to carefully
confine the solution space of SM, they can only be considered as a complementary stage for other attacks.

Prior art~\cite{rajendran13_split,wang16_sm,sengupta17_SM_ICCAD,wang17,magana16,magana17,feng17,wang17_FEOL} proposes different measures to render FEOL layouts resilient against proximity attacks.
For example, Wang \emph{et al.}~\cite{wang16_sm} as well as Sengupta \emph{et al.}~\cite{sengupta17_SM_ICCAD} perturb the placement of
gates.
However, in~\cite{wang16_sm,sengupta17_SM_ICCAD} it has been shown that splitting after higher layers---which is essential to limit the commercial cost of SM~\cite{xiao15}---can
undermine the protection.
Interestingly,
to some degree, this even holds true when layout randomization is
applied~\cite{sengupta17_SM_ICCAD}.
Such limitation of placement-centric schemes is due to the fact that any placement perturbation is eventually resolved by routing.

Routing-centric protection schemes like those in~\cite{rajendran13_split,wang17,magana16,feng17,magana17} are typically post-processing the original layouts and thus subject to constraints in routing resources and PPA budgets.
Hence, such schemes can be limited to relatively
few and/or short wiring detours, which may be easy to attack.
For example, 
Rajendran \emph{et al.}~\cite{rajendran13_split} propose to swap the pins of IP modules and reroute them to mislead an attacker.
Since the related perturbations only cover the system-level interconnects, this scheme ($i$) cannot protect against gate-level IP piracy and ($ii$) imposes a relatively small
solution space for the attacker.
As reported in~\cite{rajendran13_split} itself, on average 87\% of the connections can still be recovered.
Besides, enforcing routing detours in the BEOL requires customizing the design flow, which itself can be challenging.  For example, the schemes in
\cite{magana16,magana17} only allow
for implicit detours, namely by inserting routing blockages.

\section{Our Concept (Under Attack)}
\label{sec:concept}

\begin{figure}[tb]
\centering
\includegraphics[width=.82\columnwidth]{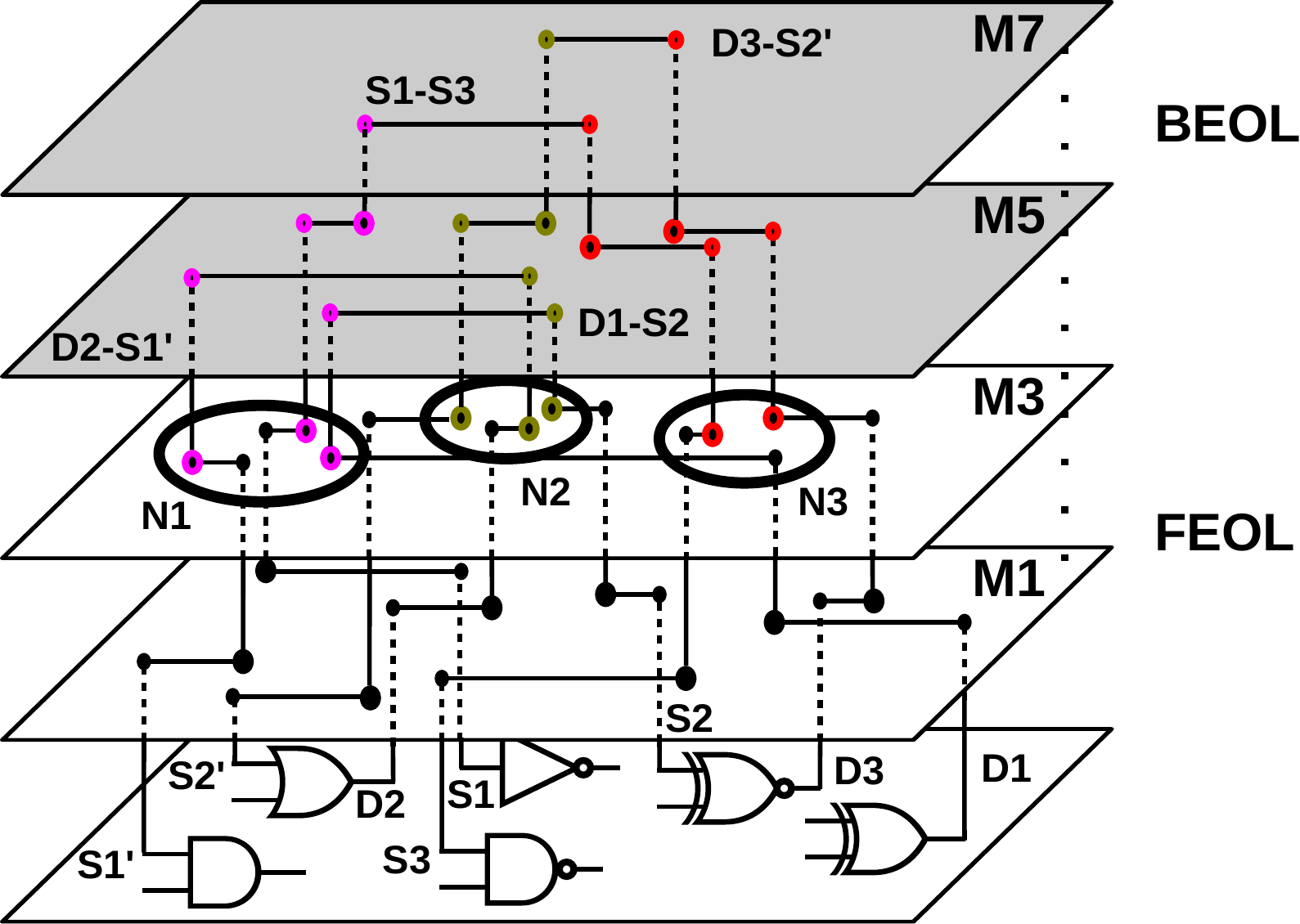}
\smallerspacecaption
\caption{
	The original netlist is randomized, to N1 (D1 driving S1 and S1'), N2, etc. We place and route the erroneous design, resulting in a
		FEOL layout which is misleading regarding both placement and routing.  Next, we restore the true functionality through the
		BEOL by rerouting; now D1 is driving S2, D2 is driving S1', etc. (The driver for sinks S1 and S3 is not illustrated in this
				conceptional figure.)
\label{fig:concept}
}
\smallerspacecaption
\end{figure}

Unlike most prior art which manipulates the placement and/or routing at the layout level in a post-processing manner, our concept targets on
the netlist itself---namely by partially randomizing it.
This helps us retain the misleading modifications throughout any regular design flow, thereby
obtaining more resilient FEOL layouts.
These layouts are corrected only in the BEOL
(Fig.~\ref{fig:concept}).
Next, we outline how our concept misleads state-of-the-art attack schemes.
We confirm our intuition in Sec.~\ref{sec:security}, where we report on superior values for key security metrics.

Any attack---even when perfect recovery rates are presumed---is bound to observe logic and timing paths of the erroneous netlist.
Regarding the attack proposed by Wang \emph{et al.}~\cite{wang16_sm}, the physical proximity of gates, the load and timing constraints, as well as the direction of dangling wires are
all capturing the erroneous netlist, not the original one.
For example, a large buffer such as \emph{BUFX8} typically hints 
that its sink(s) is/are relatively far away.
In the original netlist, however, this buffer may actually drive some nearby sink(s).
As for the routing-centric attack in~\cite{magana16},
	we note that
randomizing the netlist helps to enlarge the solution space significantly 
when compared to the original layouts (and even when compared to \emph{naive lifting},
Sec.~\ref{sec:security}).
This will naturally hinder any subsequent attacks (again, which are yet to be demonstrated).

We shall assume that the attacker knows the principle of our protection scheme;
hence she/he expects misleading FEOL layouts.
Without the
BEOL disclosed to her/him, however, a naive attacker can arguably only resort to
brute-force, i.e., enumerating all possible netlists, which is computationally prohibitive (Sec.~\ref{sec:background}).
A more sophisticated attacker may seek to exclude ``unreasonable'' logic and timing paths arising due to randomization.
Doing so, however, is subject to ($i$) significant experience on layout design, ($ii$) the size and the (yet unknown) scope of the original netlist,
	and ($iii$) the ``degree of unreasonableness'' of paths, which is random.
We believe that such attacks are an open but interesting challenge.

\section{Methodology}
\label{sec:method}

\begin{figure}[tb]
\centering
\includegraphics[width=.9\columnwidth]{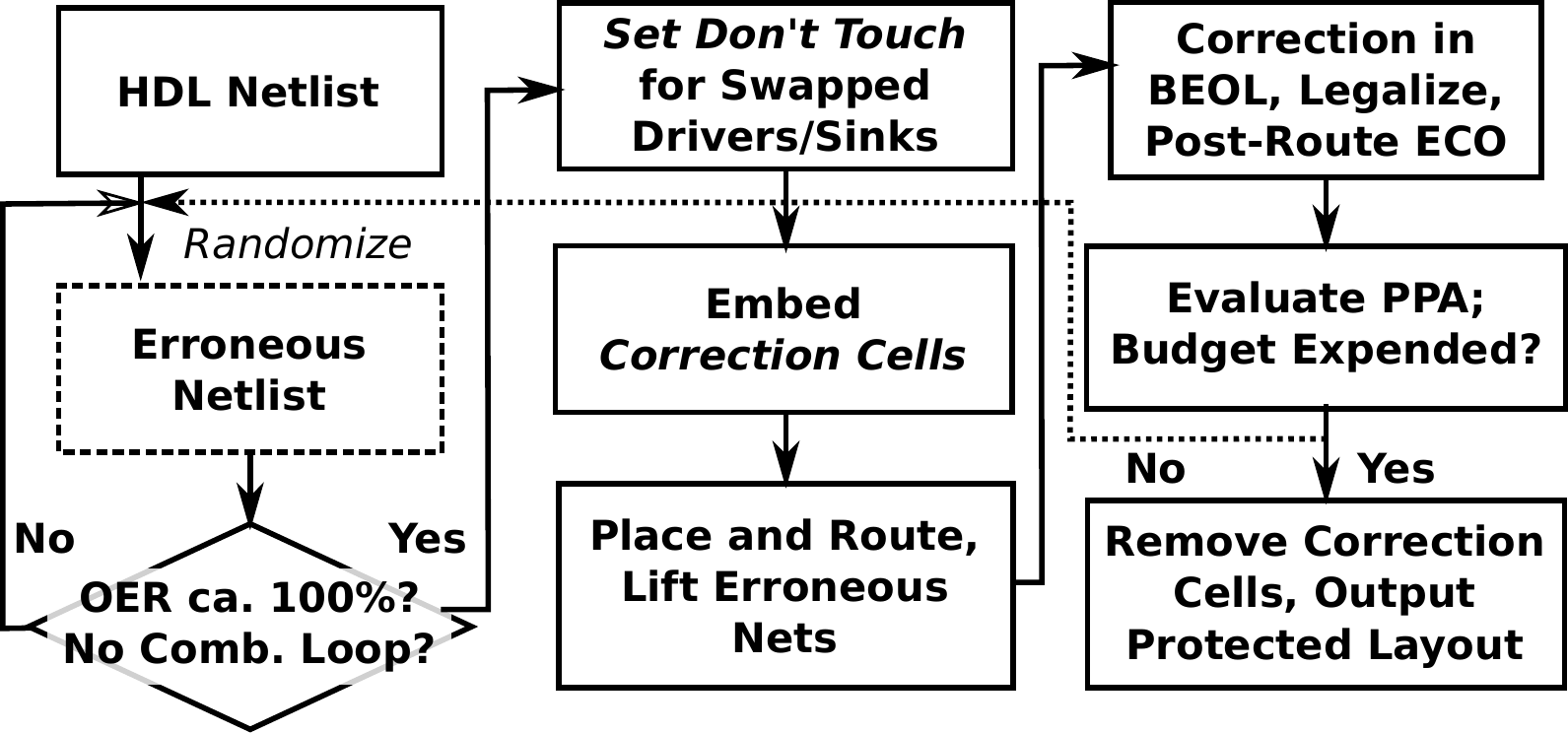}
\smallerspacecaption
\caption{The flow of our protection scheme.
\label{fig:flow}
}
\smallerspacecaption
\end{figure}

Our methodology is implemented as an extension to \emph{Cadence Innovus} with custom in-house scripts and library customization.
The steps can be summarized as follows: we ($i$) randomize the netlist, ($ii$) place and route the erroneous and misleading netlist, and ($iii$) restore the true functionality
by re-routing in the BEOL (Fig.~\ref{fig:flow}).

Next, we provide some details.
For ($i$), we iteratively randomize the netlist by swapping the connectivity between randomly selected pairs of drivers and their sinks.\footnote{In case
	the netlist imposes some alignment constraints, e.g., for datapaths, the related gates have to be ignored for randomization.}
While doing so, we ensure that no combinatorial loops arise in the modified netlist by any of the random swaps---loops would help an
attacker to identify those modifications~\cite{wang16_sm}.
We perform swapping until the OER approaches 100\%, which means that the modified netlist will induce some errors for any input.
We also keep track of the original connectivity and the swapped drivers/sinks.
For ($ii$), the randomized netlist is loaded into \emph{Cadence Innovus}. Initially, the swapped drivers/sinks are marked as \emph{do not touch} to avoid logic
restructuring/removal of the related nets.
The netlist is then placed and optimized for timing, power, and congestion.
Before routing, the nets connecting the swapped drivers/sinks are prepared for lifting to M6 (or M8) with the help of customized \emph{correction cells}.
Note that these correction cells are not impacting the FEOL layout; their scope is lifting and correction of nets in the BEOL (see also below).
Next, the design is placed and routed again in \emph{ECO} mode, to implement lifting of the swapped nets.
For ($iii$), the true connectivity is restored in the BEOL with the help of the correction cells and the tracked original connectivity.
The design is rerouted and taken through \emph{postRoute} optimization to improve timing and resolve DRC issues.
In case the PPA budget is not expended yet, we repeat the steps to introduce more randomization. Otherwise,
we remove the correction cells, and export DEF/Verilog files for further layout/security analysis.

\begin{figure}[tb]
\centering
\smallerspacecaption
\smallerspacecaption
\subfloat[]{\includegraphics[width=.39\columnwidth]{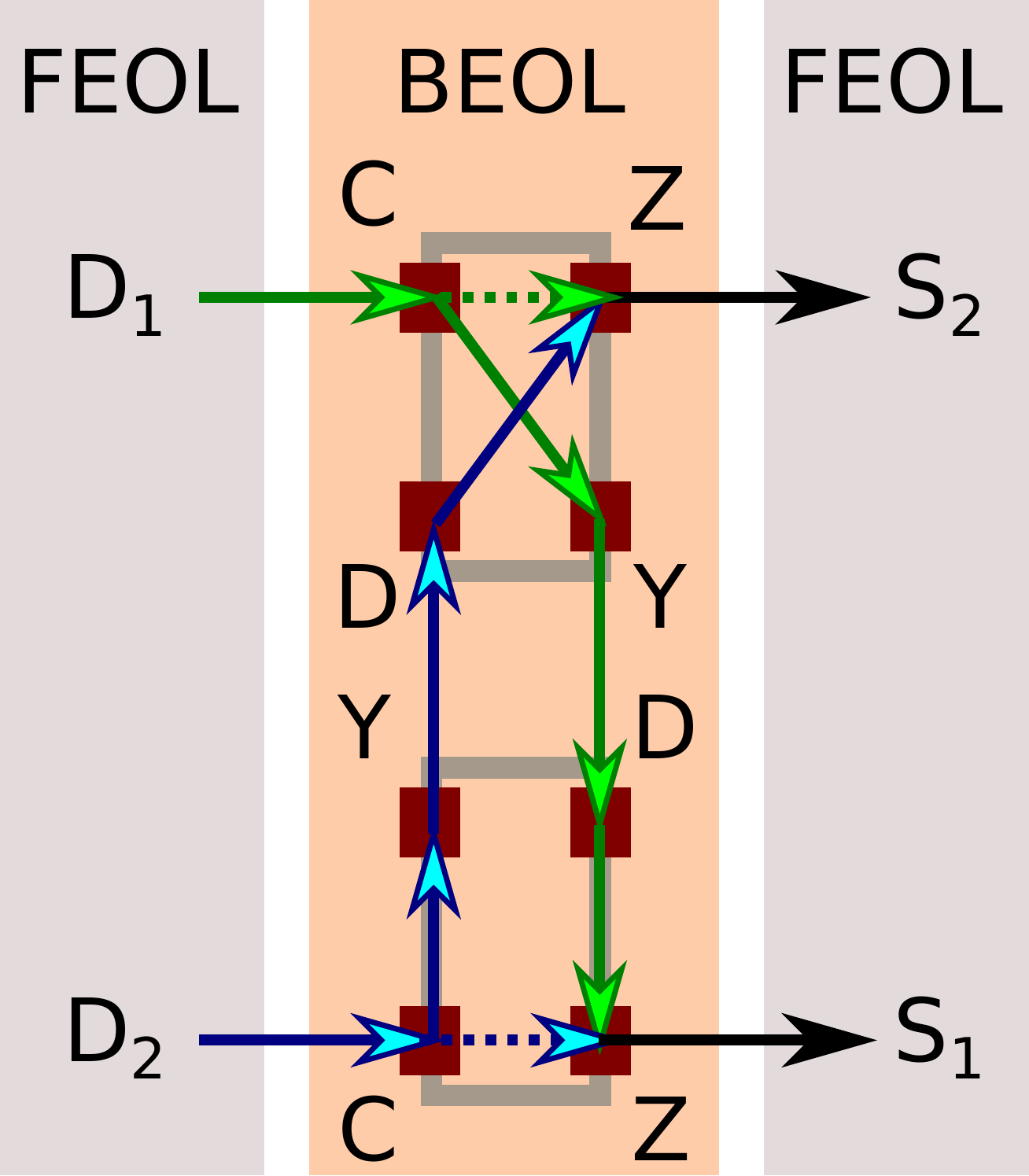}}\qquad\quad
\subfloat[]{\includegraphics[width=.345\columnwidth]{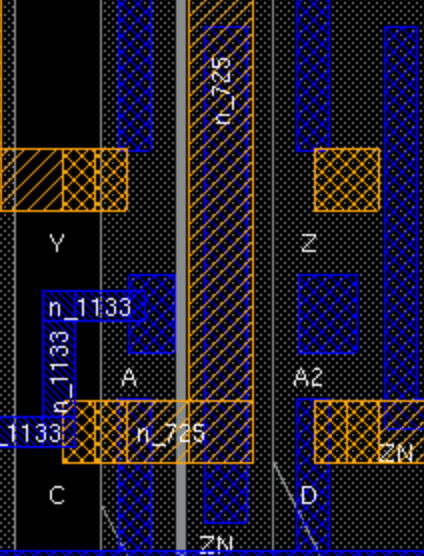}}
\smallerspacecaption
\caption{Our correction cells, in conceptional view (a) and layout view (b).
	For (a), dashed arrows indicate the misleading arcs for initial place and route,
	whereas regular arrows represent the true paths implemented later by re-routing.
		It is important to note that re-routing is always between pairs of correction cells.
		In (b), a correction cell with its pins in M6 is seen overlapping with an inverter (pins in M1).
}\label{fig:cell}
\smallerspacecaption
\smallerspacecaption
\end{figure}

Key points for the physical design of \emph{correction cells} are given next.
We provide our cell implementation on top of the \emph{Nangate 45nm} library in~\cite{webinterface}.
Figure~\ref{fig:cell} illustrates
our
cells.

\begin{itemize*}

\item The correction cells are modeled as 2-input-2-output OR gates; C, D are input pins, and Y, Z are output pins.

\item There are four possible \emph{arcs} (C to Y, C to Z, D to Y, and D to Z).
The arc C to Z is used to implement the erroneous netlist during initial place and route. When restoring the true functionality, the arcs C to Z and D to Y are disabled
(\emph{set\_disable\_timing}) so that only true paths are considered for proper timing and power optimization and evaluation.

\item
All pins are set up in a higher metal layer (here M6 or M8) to enable lifting and routing of wires in the
BEOL.
The dimensions and offsets for the pins are chosen such that they can be placed onto the tracks of the respective metal layers---this helps to minimize the routing congestion. 

\item The correction cells can freely overlap with standard cells.
That is because standard cells have their pins exclusively in lower metal layers,
whereas correction cells neither impact those layers nor the device layer.
We implement custom legalization scripts accordingly.
These scripts further prevent
different correction cells from overlapping with each other.

\item The buffer cell \emph{BUFX2} is leveraged for power and timing characteristics for the correction cells.
We can refrain from detailed library characterization since the correction cells only implement some BEOL wires.

\item To enable proper ECO optimization, the correction cells are set up for load annotation at design time.
That is required to capture the capacitive loads of ($i$) the wires running from the
correction cells to the sinks and ($ii$) the sinks themselves.

\end{itemize*}

It is important to note that re-routing (to restore the original netlist) is always between pairs of correction cells, not only within one cell.
This implies that an attacker may know the scheme of correction cells (i.e., true paths are between C and Y and between D and Z), but she/he cannot
derive the original netlist from that knowledge alone. Instead, she/he has to identify the correct pairs of cells, which is
hampered by two facts: ($i$) retracing the BEOL-centric correction cells in the FEOL is challenging, and even when the attacker
succeeds, then ($ii$) the distances between correct pairs of cells are randomized, based on the erroneous netlist being placed and
routed.
For ($ii$), recall that this misleads any state-of-the-art proximity attack to begin with (Sec.~\ref{sec:concept}).
For ($i$), note that the pins of correction cells are in higher layers (M6 or M8) whereas the layout is split 
after lower layers, necessitating some wiring paths in between.
Hence, the dangling wires related to the correction cells are unlikely to be as distinct as in Fig.~\ref{fig:cell}(b) but rather
spread out.

\section{Results}
\label{sec:results}

\subsection{Experimental Setup}
\label{sec:setup}

\textbf{Test cases:}
We evaluate our proposed defense on 12 benchmarks, seven from the \emph{ISCAS-85} suite and five from the industrial \emph{IBM superblue} suite~\cite{viswanathan11}. 
We convert the \emph{superblue} benchmarks
(initially defined in \emph{Bookshelf} format)
to \emph{Verilog} files 
using scripts from~\cite{kahng14}.

\textbf{Setup for layout evaluation:}
Our techniques are implemented for \emph{Cadence Innovus 16.15} using custom in-house \emph{TCL} scripts, which impose negligible runtime
overheads.
    We leverage the \emph{Nangate 45nm Open Cell Library}~\cite{nangate11} with ten metal layers.  
Correction cells are set up in M6 for \emph{ISCAS-85} and in M8 for \emph{superblue} benchmarks.
Conservative PPA analysis is carried out for the slow process corner and a supply voltage of 0.95V.
We ensure that all layouts are free of congestion by choosing appropriate utilization rates.
We allow PPA budgets of 20\% for \emph{ISCAS-85} and 5\% for \emph{superblue} benchmarks.

\textbf{Setup for security evaluation:}
In line with the prior art,
   we assume that the attacker has access to the FEOL layout and the technology libraries, but
she/he cannot access working chips yet to be manufactured. 
We utilize the network-flow attack~\cite{wang16_sm} for \emph{ISCAS-85} benchmarks and the routing-based attack \emph{crouting}~\cite{magana16} for \emph{superblue}
benchmarks.\footnote{
	For the latter, note that
the advanced attack as proposed by Maga\~{n}a \emph{et al.} in~\cite{magana17} has not been available to us at the time of writing.
Also, note that we have to split the DEF files obtained by \emph{Cadence Innovus} and convert them to \emph{.rt/.out} files using custom
scripts~\cite{webinterface}, as the scripts provided by Maga\~{n}a \emph{et al.}
are tailored for academic routers.}
The OER and HD are computed using
\emph{Synopsys VCS} upon applying 1,000,000 test patterns for each netlist; functional equivalence is validated using \emph{Synopsys Formality}.

\textbf{Comparative study:}
Besides the \emph{correction cells}, we also implement another set of custom cells, called \emph{naive lifting cells}. These cells are implementing the same principle of lifting
wires,
but without inducing erroneous connections. We apply these cells using our flow on original layouts to obtain a baseline called \emph{naive lifting}.

The protected layouts of~\cite{wang16_sm, wang17} have been made available to us as DEF files. 
Since there are no definite indications of the split layer in the respective publications,
we average the security metrics (CCR, OER, and HD) for splitting after layers M3, M4, and M5 respectively. 

\textbf{Open source:}
   We make our \emph{correction cells} and \emph{naive lifting cells}
available to
the community, along with the protected layouts, in~\cite{webinterface}.
We also provide our DEF splitting and conversion script.

\subsection{Security Evaluation}
\label{sec:security}

\begin{figure*}[tb]
\centering
\smallerspacecaption
\smallerspacecaption
\smallerspacecaption
\subfloat[]{\includegraphics[width=.31\textwidth]{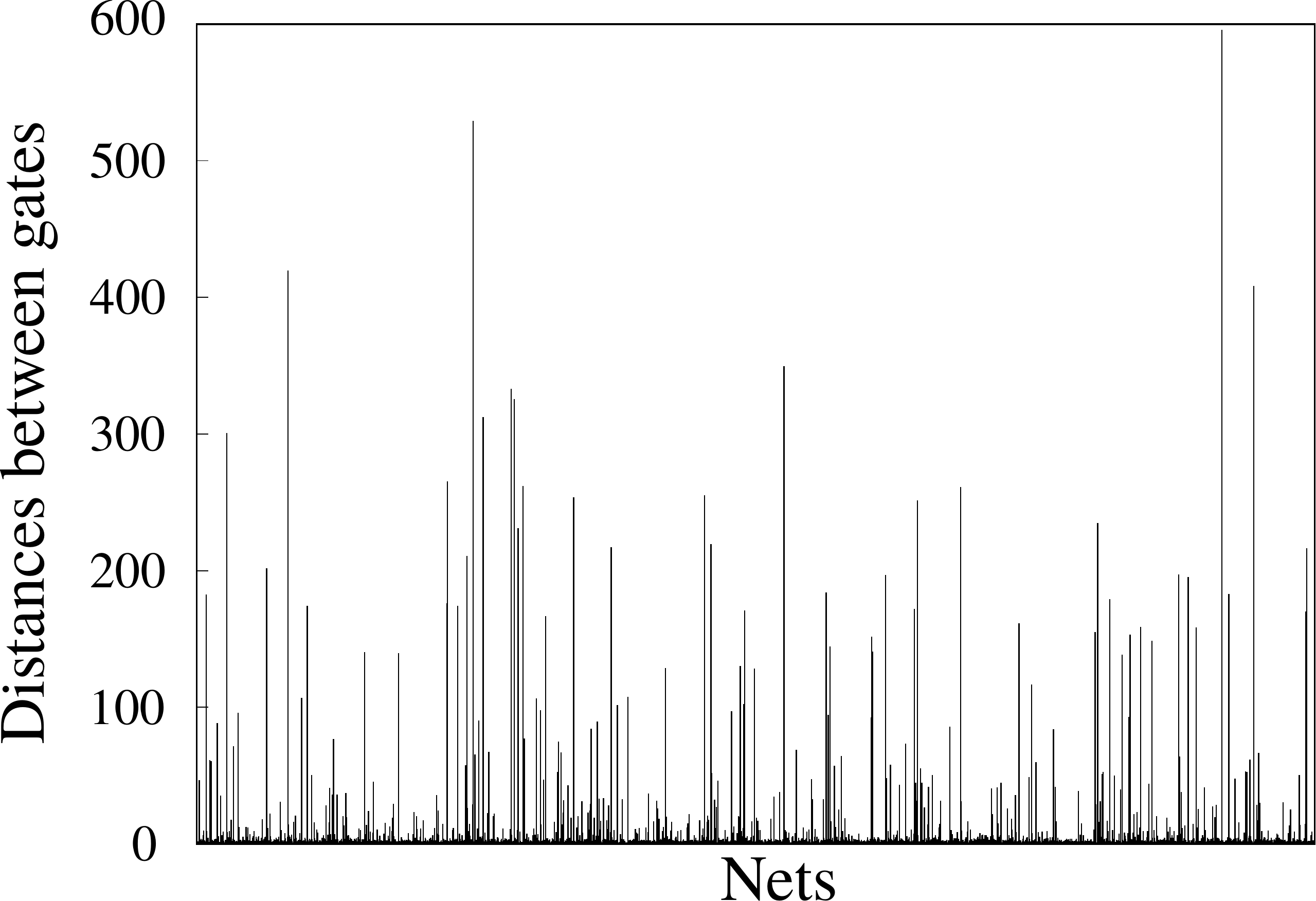}}\hfill
\subfloat[]{\includegraphics[width=.31\textwidth]{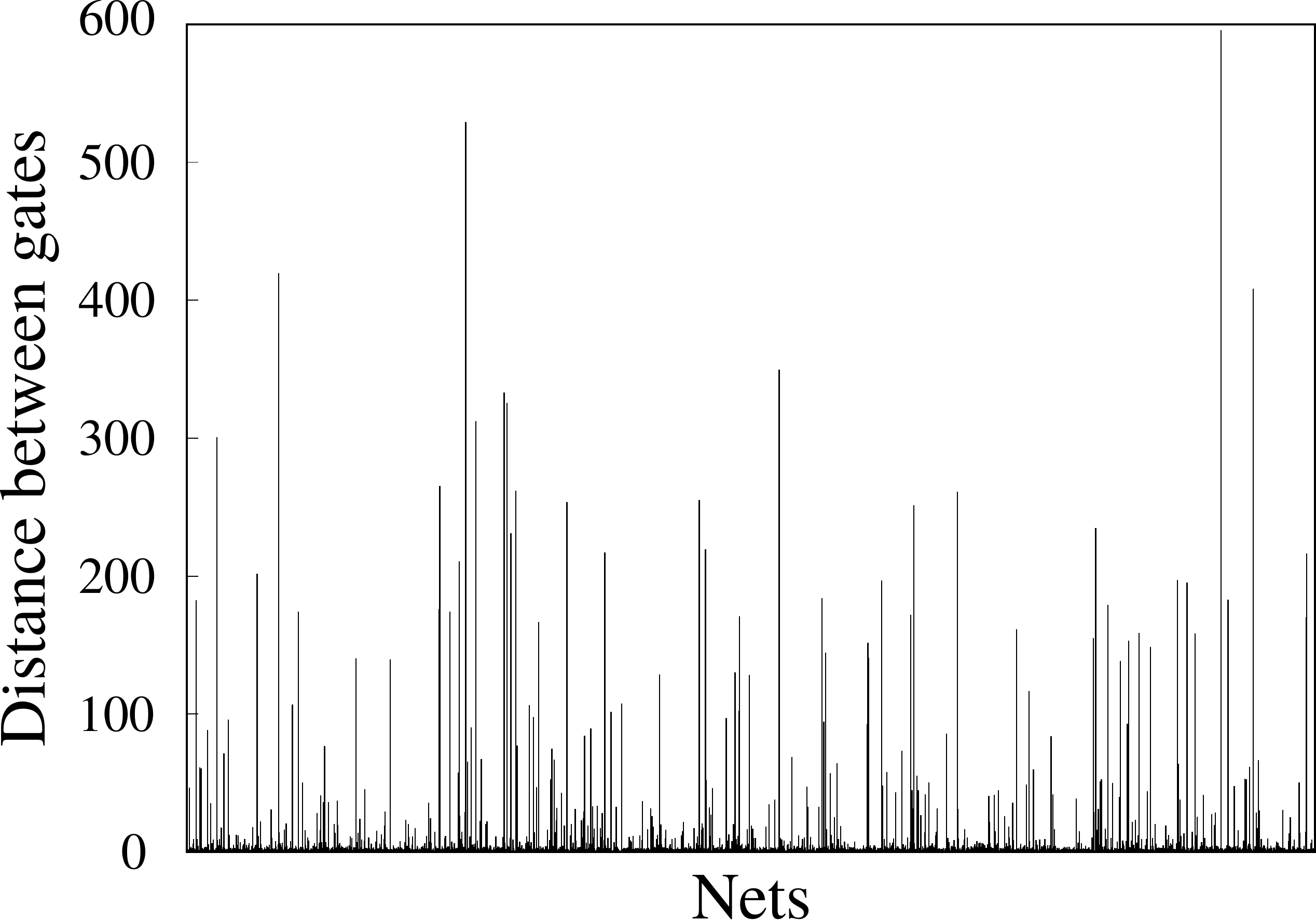}}\hfill
\subfloat[]{\includegraphics[width=.31\textwidth]{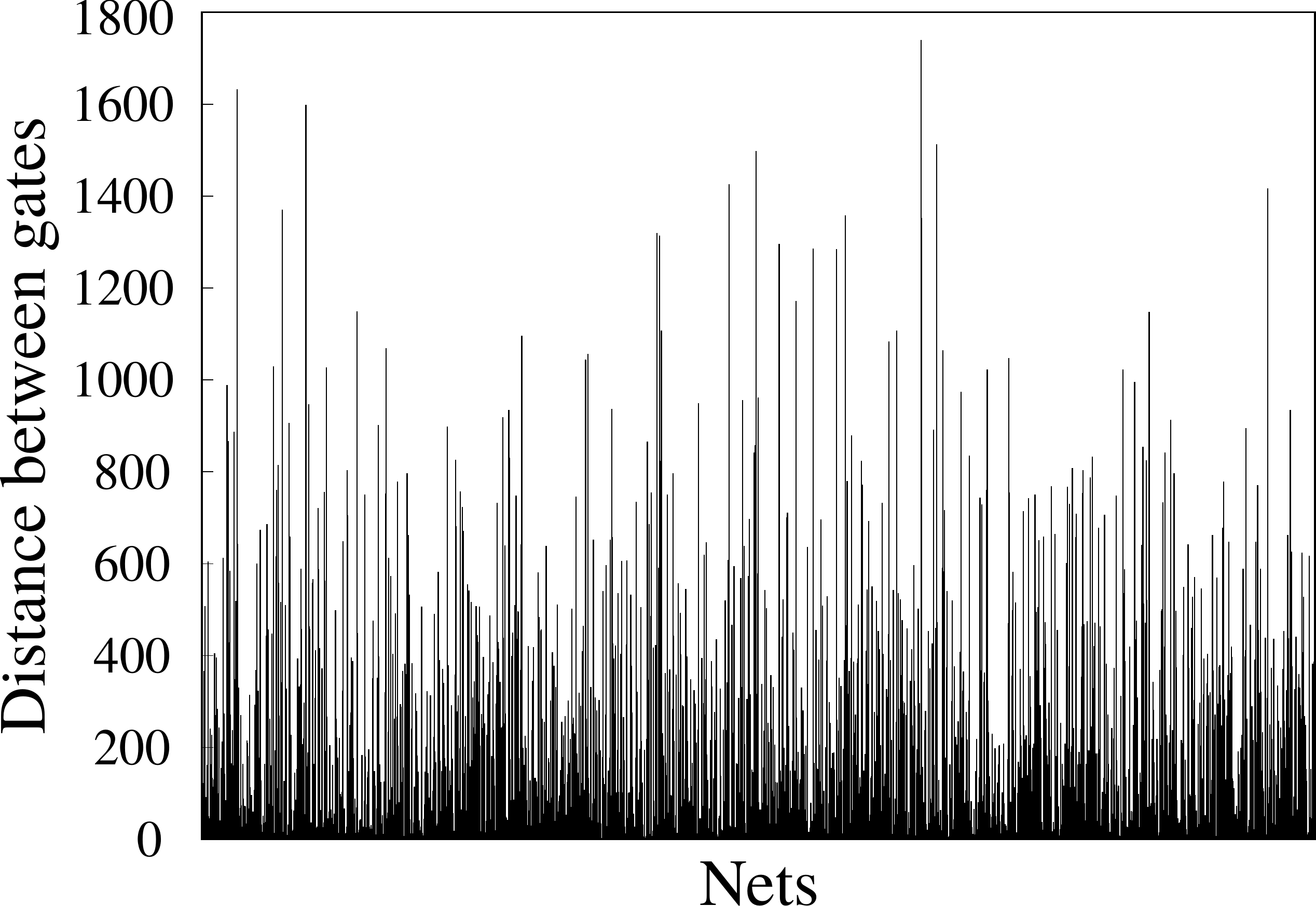}}
\smallerspacecaption
\smallerspacecaption
\caption{
Distances between drivers/sinks for original (a), naively lifted (b), and our layouts (c), for
	\emph{superblue18} benchmark.
\label{fig:SB18_distances}
}
\end{figure*}

\textbf{Protection of the placement}: Recall that our scheme is based on randomly modified netlists, leading to erroneous FEOL layouts which
	are corrected only in the BEOL.
It is intuitive that
the distances of \emph{truly} connected gates will be randomly distributed, thereby misleading proximity attacks. 
In fact,
we achieve a significant increase of the distances along with a widely varied distribution
(Fig.~\ref{fig:SB18_distances} and Table~\ref{tab:layout-distances}).
This finding is corroborated by
the superior
resilience for \emph{ISCAS-85} benchmarks
(Tables~\ref{tab:sec-comp1} and \ref{tab:sec-comp2});
see further below for a comparative study on placement protection.

\begin{figure}[tb]
\centering
\smallerspacecaption
\smallerspacecaption
\includegraphics[width=\columnwidth]{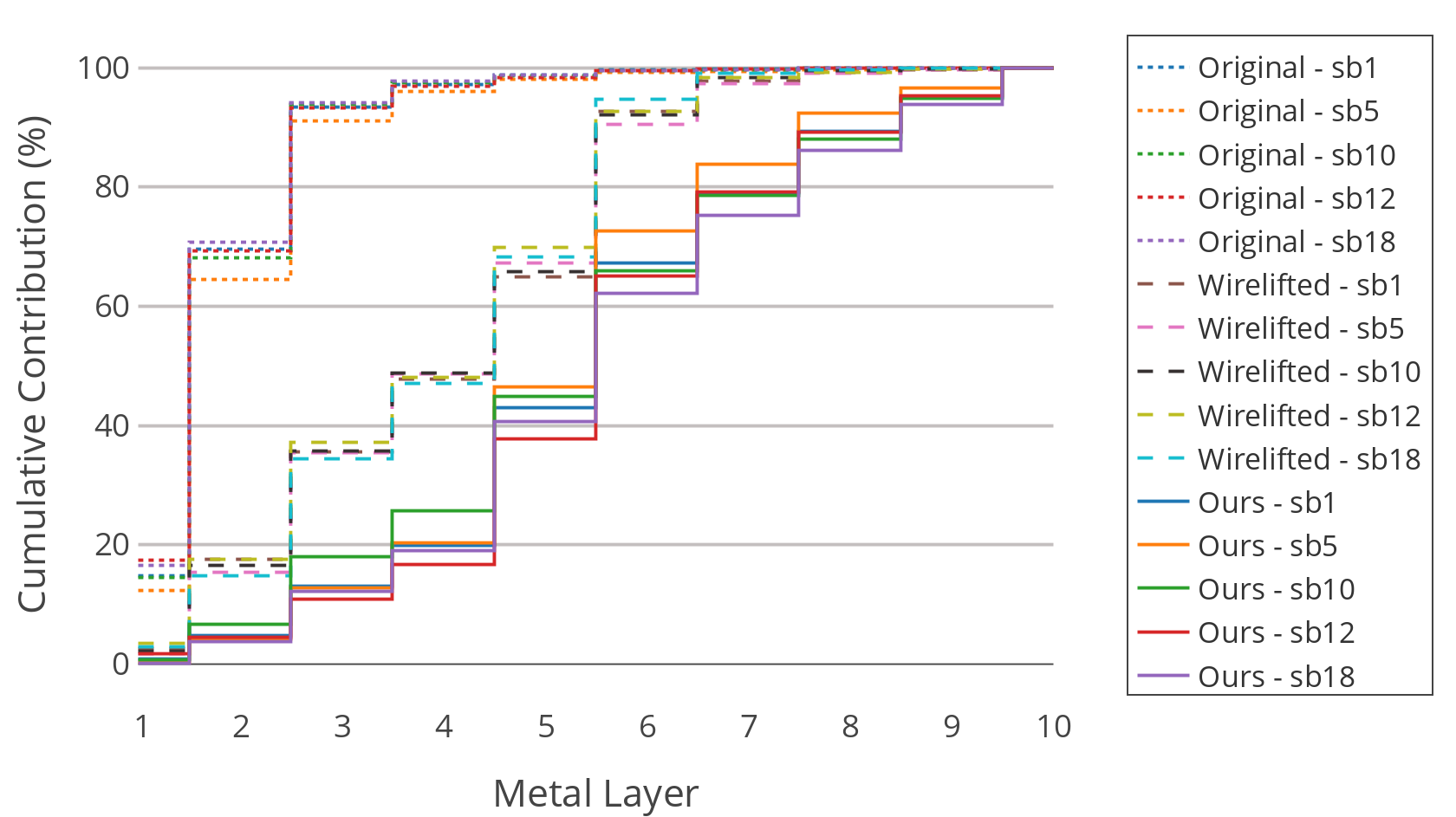}
\smallerspacecaption
\smallerspacecaption
\smallerspacecaption
\caption{Contribution of the metal layers to wirelength for randomized nets in \emph{superblue} benchmarks.
\label{fig:metal_sb}
}
\smallerspacecaption
\smallerspacecaption
\end{figure}

\textbf{Protection of the routing}:
Recall that placement-centric protection schemes are offset by routing, especially once splitting is conducted after higher metal layers
(Sec.~\ref{sec:background}),
rendering routing-centric schemes more promising in that context.
Figure~\ref{fig:metal_sb} contrasts the contribution of each metal layer towards the wirelength
		for \emph{superblue} benchmarks.
For original layouts, the majority of wiring is found in the lower metal layers which provides significant leverage for an attacker.
As for naively lifted layouts, the wiring is more evenly distributed across the metal layers which \emph{may} help to protect the layouts.
However, it is important to note that naive lifting
cannot help dissolve the distances between connected gates (Fig.~\ref{fig:SB18_distances}).
In contrast, our scheme holds the majority of wiring in the higher layers \emph{and} dissolves the true connectivity in the FEOL, offering less leverage for an
attacker \emph{and} enabling splits after higher layers.

\begin{table}[tb]
\centering
\footnotesize
\caption{
Distances between connected gates (in microns).
}
\smallerspacecaption
\input{incl/tab-layout-distances}
\label{tab:layout-distances}
\smallerspacecaption
\end{table}

We observe that our scheme is significantly more effective than naive lifting in terms of increasing vias/vpins in higher layers
(Table~\ref{tab:layout-vias}). Additional vias/vpins
induce more nets in the BEOL, rendering
routing-centric attacks more challenging.
For example, taking M5 as the split layer, our scheme increases the vias V56 by 30.65\% on average when compared to naive lifting.
We observe accordingly that the \emph{crouting} attack~\cite{magana16} is impaired by larger lists of candidates and more vpins (Table~\ref{tab:sec-routing}).
Recall that even seemingly small increases in those metrics imply a large-scale, polynomial
increase of complexity---our scheme increases both metrics compared to original as well as lifted layouts.

\begin{table*}[tb]
\centering
\footnotesize
\setlength{\tabcolsep}{1.5mm}
\caption{
Comparison of additional vias for naively lifted layouts and our proposed scheme over original \emph{IBM superblue} layouts. For a fair comparison, we randomize the same set of
	nets. We ensure zero die-area overhead and all layouts are DRC-clean.
}
\smallerspacecaption
\input{incl/tab-layout-vias}
\label{tab:layout-vias}
\end{table*}

\begin{table}[tb]
\centering
\smallerspacecaption
\footnotesize
\caption{Results for \emph{crouting} attack~\cite{magana16}.
Comparison of vpins and candidate list size (E[LS]) as a function of bounding box.
	N/A denotes attack failures.
}
\smallerspacecaption
\input{incl/tab-sec-routing}
\label{tab:sec-routing}
\smallerspacecaption
\smallerspacecaption
\end{table}

In short, our scheme ($i$) keeps the major share of wirelength
in the BEOL
		(Fig.~\ref{fig:metal_sb})
and ($ii$) significantly increase the via counts (Table~\ref{tab:layout-vias}), all while inducing misleading routing in the FEOL.

\textbf{Comparison with the prior art:}
We contrast the resilience of our scheme against various prior art in Tables~\ref{tab:sec-comp1} and \ref{tab:sec-comp2}.
As expected, the original layouts are most prone to proximity attacks:
on average 94\% CCR and 7\% HD can be achieved when running the attack of~\cite{wang16_sm}.
Selective gate-level placement perturbation as proposed in~\cite{wang16_sm} offers only a marginal improvement over unprotected layouts, with an attacker making 92\% correct
connections and experiencing 15\% HD.
Sengupta \emph{et al.}~\cite{sengupta17_SM_ICCAD} proposed four different protection strategies, 
and while the CCR is reduced to 63\% on average, it should be noted that those techniques can become impractical to protect larger designs (in terms of excessive PPA overheads).
Swapping of block pins as proposed in~\cite{rajendran13_split}
is also limited; the attacker can still correctly infer 87\% of the missing system-level interconnects.
The
scheme proposed by Wang \emph{et al.}~\cite{wang17} reduces the CCR to about 72\%.
More recently, Feng \emph{et al.}~\cite{feng17} proposed a routing-based scheme
which can reduce the CCR
significantly to about 21\%.

Our scheme offers the best protection as it can reduce the CCR to the ideal value of 0\%---none of the nets randomized in the FEOL are correctly inferred by an attacker.
We attribute this superior result to the holistic mitigation of placement \emph{and} routing hints.
Besides CCR, our scheme also achieves an OER of $\approx$100\% (only~\cite{wang17} reports similar numbers).
Finally, our average HD is 40.4\% which is another significant improvement over the prior art.

Besides the \emph{crouting} attack, Maga\~{n}a \emph{et al.}~\cite{magana16,magana17} also proposed routing-centric protection schemes.
As discussed, a key metric for their evaluation is the number of vias/vpins.
In Table~\ref{tab:sec-comp-magana17}, we compare their most recent results~\cite{magana17} with ours, whereas we set up the \emph{correction cells} for lifting wires to M8.
Since our scheme increases the via count in those higher layers to a much larger degree, we believe that it is more resilient than that of
Maga\~{n}a \emph{et al.}\footnote{Recall that the attacks in~\cite{magana16,magana17} do not provide actual netlists, hence we cannot compare for related metrics such as CCR, OER, and HD.
Besides, note that Maga\~{n}a \emph{et al.}~\cite{magana17} cautioned on wire lifting, since routers may be misguided by lifting which, in turn, can impact PPA. Our scheme
	exhibits reasonable PPA cost despite lifting (Sec.~\ref{sec:layout}).}

\begin{table*}[tb]
\centering
\footnotesize
\smallerspacecaption
\caption{
Comparison with placement perturbation schemes.
	The metrics for our scheme are averaged for splitting after M3, M4, and M5; the metrics for the prior art are quoted.
	All values are in percentage. The attacks are based on~\cite{wang16_sm}. 
}
\smallerspacecaption
\input{incl/tab-sec-comp1}
\label{tab:sec-comp1}
\smallerspacecaption
\end{table*}

\begin{table*}[tb]
\centering
\footnotesize
\caption{
Comparison with routing perturbation schemes. The setup is the same as in Table~\ref{tab:sec-comp1}.
}
\smallerspacecaption
\input{incl/tab-sec-comp2}
\label{tab:sec-comp2}
\smallerspacecaption
\end{table*}

\begin{table}[tb]
\centering
\footnotesize
\caption{Comparison with~\cite{magana17} w.r.t.~additional via count. Layouts are split after M6 and true connectivity restored in M8.
}
\smallerspacecaption
\input{incl/tab-sec-comp-magana17}
\label{tab:sec-comp-magana17}
\end{table}

\subsection{Layout Evaluation}
\label{sec:layout}

\begin{figure}[tb]
\centering
\smallerspacecaption
\smallerspacecaption
\includegraphics[width=.95\columnwidth]{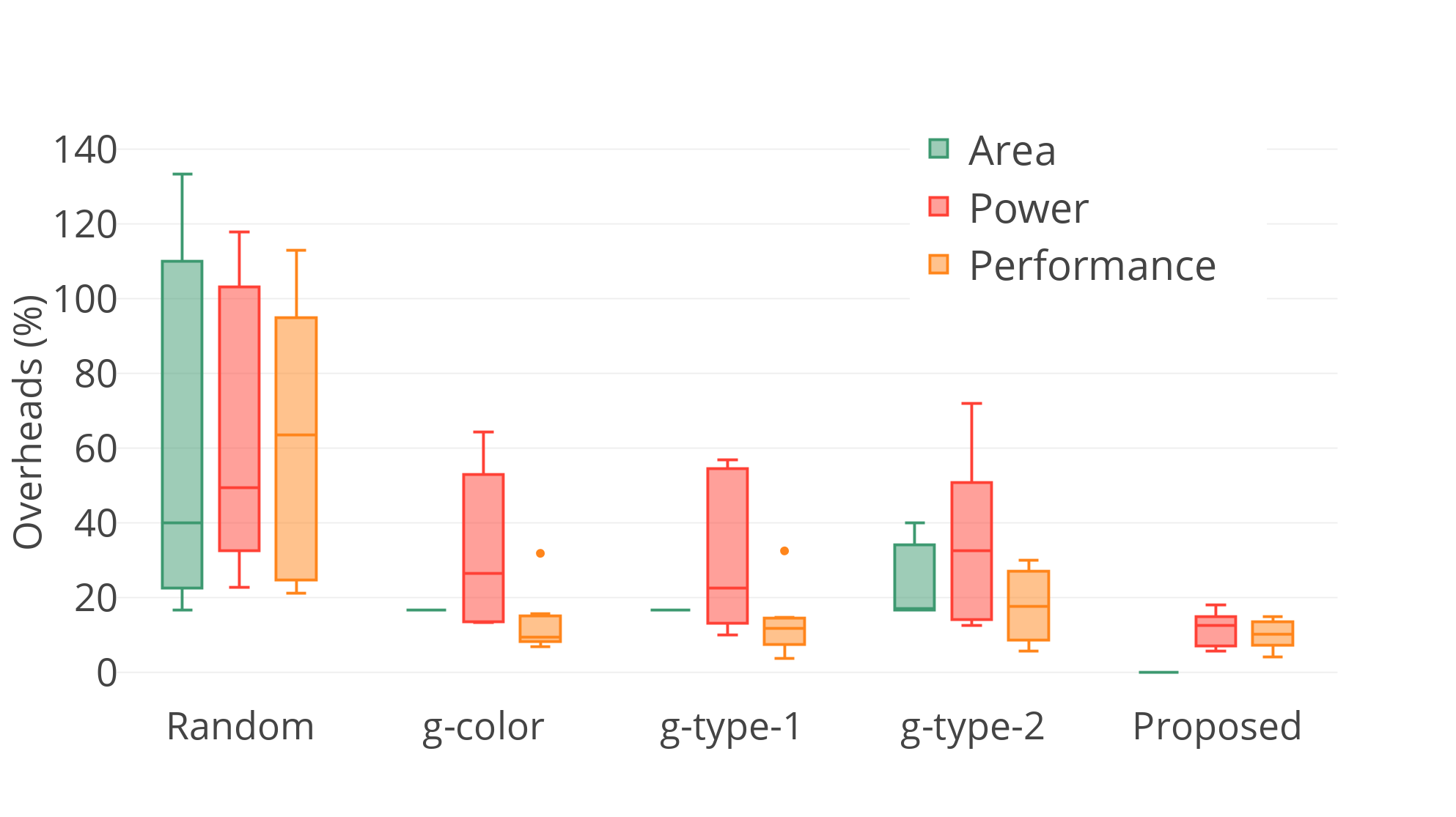}
\smallerspacecaption
\smallerspacecaption
\smallerspacecaption
\caption{Comparison with~\cite{sengupta17_SM_ICCAD} on \emph{ISCAS-85} benchmarks.
\label{fig:PPA_vs_ICCAD17}
}
\smallerspacecaption
\end{figure}

We estimate the area cost with regard to die outlines.
That is because our correction cells do not impact the device layer, but they mandate re-routing which may require larger outlines to
maintain DRC-clean layouts. In practice, however, we obtain no area cost at all (Fig.~\ref{fig:PPA_vs_ICCAD17}).

Since nets are lifted to BEOL layers (M6/M8) and rerouted,
we naturally observe some increases in wirelength. These overheads, however, translate only to some degree to power and delay cost.
On average, 
11.5\% and 10\% power and delay cost arise for the \emph{ISCAS-85} benchmarks.
For the \emph{superblue} benchmarks, average overheads are 3.5\% and 2.7\% for power and delay, respectively. We found that these overheads are 3.4\% and 2.6\%
higher than those induced for \emph{naive lifting}.
In this context, recall that our scheme outperforms naive lifting as well as prior protection schemes in terms of security.

Note that most prior studies do not report on detailed layout evaluation and PPA results.
In a recent study by Sengupta \emph{et al.}~\cite{sengupta17_SM_ICCAD}, however, the authors report on PPA results for their protection scheme. We contrast their results with ours
in Fig.~\ref{fig:PPA_vs_ICCAD17}; our scheme induces on average lower overheads on all area, power, and delay.

\section{Conclusion}
\label{sec:conclusion}

Multiple studies recently questioned the security of split manufacturing.
In this work, we raise the designer's game to protect against malicious FEOL parties.
Our idea is to randomize the functionality and connectivity in the netlist, place and route that misleading design, and restore the true functionality only through the BEOL.

We contrast our scheme to recent state-of-the-art defense techniques, while leveraging both placement- and routing-centric attacks. We observe that our protected layouts are
significantly more resilient while exhibiting reasonable PPA cost, even on large-scale industrial benchmarks.
Another contribution in our scheme is that we readily support splitting after higher layers (e.g., after M6), thereby limiting commercial cost.
In future work, we seek to protect against further threats such as insertion of hardware Trojans
and to formulate strategies towards ``provably secure SM.''

\section*{Acknowledgements}
\label{sec:acknowledgements}

The authors are grateful to Jeyavijayan (JV) Rajendran (Texas A\&M)
for providing the network-flow attack and protected layouts of~\cite{wang16_sm, wang17}.
This work was supported in part by the
	Center for Cyber Security (CCS) at
     NYU New York/Abu Dhabi
(NYU/NYUAD).

\newcommand{\BIBdecl}{\setlength{\itemsep}{-0.1em}}
\bibliographystyle{IEEEtran}
\bibliography{main}

\end{document}

%% file: abstract.tex
	Split manufacturing (SM) seeks to
	protect against piracy of intellectual property (IP) in chip designs.
	Here we propose a scheme to manipulate both placement and routing in an intertwined manner, thereby increasing the resilience of SM layouts.
	Key stages of our scheme are to (partially) randomize a design, place and route the erroneous netlist,
	and restore the original design by re-routing the BEOL.
	Based on
	state-of-the-art proximity attacks,
	we demonstrate that our scheme notably excels over the prior art (i.e., 0\% correct connection rates).
	Our scheme induces controllable PPA overheads and lowers commercial cost (the latter by splitting at higher layers).

%% file: incl/tab-layout-distances.tex
\begin{tabular}{|*{5}{c|}}
\hline
\textbf{Benchmark}
& \textbf{\emph{Layout}}
& \textbf{Mean}
& \textbf{Median}
& \textbf{Std. Dev.}
\\ \hline

\multirow{3}{*}{\emph{superblue1}}
& \textbf{\emph{Original}} &
14.31 & 2.85 & 54.84
\\ \cline{2-5}
& \textbf{\emph{Lifted}} &
14.37 & 2.92 & 54.83
\\ \cline{2-5}
& \textbf{\emph{Proposed}} &
198.46 & 48.41 & 318.88
\\ \hline

\multirow{3}{*}{\emph{superblue5}}
& \textbf{\emph{Original}} &
14.38 & 2.99 & 49.16
\\ \cline{2-5}
& \textbf{\emph{Lifted}} &
14.39 & 2.99 & 49.17
\\ \cline{2-5}
& \textbf{\emph{Proposed}} &
244.73 & 96.9 & 328.84
\\ \hline

\multirow{3}{*}{\emph{superblue10}}
& \textbf{\emph{Original}} &
12.66 & 2.73 & 49.59
\\ \cline{2-5}
& \textbf{\emph{Lifted}} &
12.71 & 2.8 & 49.58
\\ \cline{2-5}
& \textbf{\emph{Proposed}} &
254.06 & 71.03 & 372.07
\\ \hline

\multirow{3}{*}{\emph{superblue12}}
& \textbf{\emph{Original}} &
19.06 & 3.18 & 75.37
\\ \cline{2-5}
& \textbf{\emph{Lifted}} &
19.08 & 3.23 & 75.37
\\ \cline{2-5}
& \textbf{\emph{Proposed}} &
263.21 & 81.28 & 395.26
\\ \hline

\multirow{3}{*}{\emph{superblue18}}
& \textbf{\emph{Original}} &
12.91 & 2.54 & 41.74
\\ \cline{2-5}
& \textbf{\emph{Lifted}} &
12.93 & 2.54 & 41.74
\\ \cline{2-5}
& \textbf{\emph{Proposed}} &
208.47 & 119.51 & 244.81
\\ \hline

\end{tabular}

%% file: incl/tab-layout-vias.tex
\begin{tabular}{|*{15}{c|}}
\hline
\textbf{Benchmark}
& \textbf{Nets}
& \textbf{I/O Pins}
& \textbf{Util.}
& \textbf{\emph{Layout}}
& \textbf{V12}
& \textbf{V23}
& \textbf{V34}
& \textbf{V45}
& \textbf{V56}
& \textbf{V67}
& \textbf{V78}
& \textbf{V89}
& \textbf{V910}
& \textbf{Total Vias}
\\ \hline

\multirow{3}{*}{\emph{superblue1}}
& \multirow{3}{*}{873,712}
& \multirow{3}{*}{8,320/13,025}
& \multirow{3}{*}{69}
& \textbf{\emph{Original}} &
3,016,748 & 2,334,923 & 664,292 & 239,550 & 170,423 &
82,762 & 56,170 & 34,164 & 16,249 & 6,615,281 
\\ \cline{5-15}
& & & & \textbf{\emph{Lifted (\%)}} &
0.10 & 0.56 & 1.29 & 2.44 & 3.45 & 3.28 & 1.69 & 0.68 & 0.37 & 0.61
\\ \cline{5-15}
& & & & \textbf{\emph{Proposed (\%)}} &
2.1 & 4.13 & 10.82 & 18.38 & 29.86 & 31.79 & 34.2 & 27.3 & 40.93 & 5.87
\\ \hline

\multirow{3}{*}{\emph{superblue5}}
& \multirow{3}{*}{754,907}
& \multirow{3}{*}{11,661/9,617}
& \multirow{3}{*}{77}
& \textbf{\emph{Original}} &
2,430,541 & 1,866,252 & 553,843 & 217,394 & 157,046 & 75,306 & 50,970 & 30,714 & 15,227 & 5,397,293
\\ \cline{5-15}
& & & & \textbf{\emph{Lifted (\%)}} &
0.1 & 0.8 & 1.8 & 3.3 & 4.9 & 5.0 & 2.4 & 1.2 & 0.7 & 0.9
\\ \cline{5-15}
& & & & \textbf{\emph{Proposed (\%)}} &
3.2 & 7.3 & 12.9 & 23.9 & 40.0 & 55.1 & 59.5 & 51.3 & 67.6 & 9.2
\\ \hline

\multirow{3}{*}{\emph{superblue10}}
& \multirow{3}{*}{1,147,401}
& \multirow{3}{*}{10,454/23,663}
& \multirow{3}{*}{75}
& \textbf{\emph{Original}} &
3,871,474 & 3,048,375 & 875,305 & 329,549 & 238,533 & 111,507 & 76,885 & 45,408 & 22,721 & 8,619,757
\\ \cline{5-15}
& & & & \textbf{\emph{Lifted (\%)}} &
0.04 & 0.49 & 1.11 & 2.11 & 3.02 & 3.18 & 1.49 & 0.53 & 0.24 & 0.52
\\ \cline{5-15}
& & & & \textbf{\emph{Proposed (\%)}} &
2.06 & 6.29 & 12.43 & 22.92 & 32.27 & 55.22 & 57.16 & 59.29 & 69.74 & 7.90
\\ \hline

\multirow{3}{*}{\emph{superblue12}}
& \multirow{3}{*}{1,520,046}
& \multirow{3}{*}{1,936/4,629}
& \multirow{3}{*}{56}
& \textbf{\emph{Original}} &
5,368,332 & 3,995,438 & 1,130,079 & 445,635 & 316,038 & 141,141 & 100,358 & 55,097 & 31,301 & 11,583,419
\\ \cline{5-15}
& & & & \textbf{\emph{Lifted (\%)}} &
0.03 & 0.16 & 0.42 & 0.87 & 1.33 & 1.37 & 0.58 & 0.31 & 0.2 & 0.2
\\ \cline{5-15}
& & & & \textbf{\emph{Proposed (\%)}} &
1.59 & 6.99 & 19.09 & 30.56 & 30.19 & 34.67 & 22.92 & 30.93 & -1.19 & 7.78
\\ \hline

\multirow{3}{*}{\emph{superblue18}}
& \multirow{3}{*}{670,323}
& \multirow{3}{*}{3,921/7,465}
& \multirow{3}{*}{67}
& \textbf{\emph{Original}} &
2,298,823 & 1,686,525 & 480,099 & 179,088 & 121,277 & 51,187 & 28,950 & 18,345 & 4,319 & 4,868,613
\\ \cline{5-15}
& & & & \textbf{\emph{Lifted (\%)}} &
0.05 & 0.55 & 1.56 & 3.69 & 5.62 & 5.82 & 3.10 & 0.72 & 1.09 & 0.73
\\ \cline{5-15}
& & & & \textbf{\emph{Proposed (\%)}} &
1.73 & 5.98 & 10.50 & 20.03 & 39.24 & 61.11 & 90.08 & 71.08 & 287.84 & 7.34
\\ \hline

\end{tabular}

%% file: incl/tab-sec-routing.tex
\begin{tabular}{|*{6}{c|}}
\hline
  \multirow{2}{*}{\textbf{Benchmark}}
& \multirow{2}{*}{\textbf{\emph{Layout}}}
& \multirow{2}{*}{\textbf{\#VPins}}
& \multicolumn{3}{|c|}{\textbf{E[LS] for Bounding Box}}
\\ \cline{4-6}
& & &
  \textbf{15}
& \textbf{30}
& \textbf{45}
\\ \hline

\multirow{3}{*}{\emph{superblue1}}
& \textbf{\emph{Original}} & 73,110 &
4.63 & 13.25 & 23.46
\\ \cline{2-6}
& \textbf{\emph{Lifted}} & 73,810 &
4.65 & 13.27 & 23.47
\\ \cline{2-6}
& \textbf{\emph{Proposed}} & 75,754 &
4.69 & 13.46 & 23.83
\\ \hline

\multirow{3}{*}{\emph{superblue5}}
& \textbf{\emph{Original}} & 67,194 &
 4.86 & 13.99 & 24.87
\\ \cline{2-6}
& \textbf{\emph{Lifted}} & 67,676 &
4.85 & 13.9 & 24.73
\\ \cline{2-6}
& \textbf{\emph{Proposed}} & N/A &
 N/A & N/A & N/A
\\ \hline

\multirow{3}{*}{\emph{superblue10}}
& \textbf{\emph{Original}} & 155,180 &
 5.05 & 14.54 & 25.75
\\ \cline{2-6}
& \textbf{\emph{Lifted}} & N/A &
 N/A & N/A & N/A
\\ \cline{2-6}
& \textbf{\emph{Proposed}} & 157,106 &
 4.88 & 14.1 & 25.07
\\ \hline

\multirow{3}{*}{\emph{superblue12}}
& \textbf{\emph{Original}} & 127,112 &
4.84 & 13.85 & 24.45
\\ \cline{2-6}
& \textbf{\emph{Lifted}} & 127,610 &
 4.83 & 13.79 & 24.35
\\ \cline{2-6}
& \textbf{\emph{Proposed}} & 165,106 &
 6.29 & 17.95 & 32.04
\\ \hline

\multirow{3}{*}{\emph{superblue18}}
& \textbf{\emph{Original}} & 50,026 &
3.76 & 10.86 & 19.17
\\ \cline{2-6}
& \textbf{\emph{Lifted}} & 51,970 &
 3.87 & 11.09 & 19.54
\\ \cline{2-6}
& \textbf{\emph{Proposed}} & 54,154 &
 4.26 & 12.22 & 21.74
\\ \hline

\end{tabular}

%% file: incl/tab-sec-comp1.tex
\begin{tabular}{|*{14}{c|}}
\hline
\multirow{3}{*}{\textbf{Benchmark}}
& \multicolumn{3}{|c|}{\textbf{Original Layout}} 
& \multicolumn{3}{|c|}{\textbf{Placement Perturbation \cite{wang16_sm}}} 
& \multicolumn{4}{|c|}{\textbf{Placement Perturbation \cite{sengupta17_SM_ICCAD}}} 
& \multicolumn{3}{|c|}{\textbf{Proposed}} \\
\cline{2-14}
& \multirow{2}{*}{\textbf{CCR}} & \multirow{2}{*}{\textbf{OER}} & \multirow{2}{*}{\textbf{HD}}
& \multirow{2}{*}{\textbf{CCR}} & \multirow{2}{*}{\textbf{OER}} & \multirow{2}{*}{\textbf{HD}}
& \textbf{Random} & \textbf{G-Color} & \textbf{G-Type1} & \textbf{G-Type2}
& \multirow{2}{*}{\textbf{CCR}} & \multirow{2}{*}{\textbf{OER}} & \multirow{2}{*}{\textbf{HD}}
\\
\cline{8-11}
& & &
& & &
& \textbf{CCR}
& \textbf{CCR}
& \textbf{CCR}
& \textbf{CCR}
& & &
\\ \hline
c432 & 92.4 & 75.4 & 23.4 &
90.7 & 98.8 & 41.8 &
68.1 & 84.4 & 89.8 & 78.8 &
0 & 99.9 & 48.4
 \\ \hline
 
c880 & 100 & 0 & 0 &
96.8 & 15.8 & 1.2 &
56.1 & 84.3 & 81.4 & 78.5 &
0 & 99.9 & 43.4
 \\ \hline
 
c1355 & 95.4 & 59.5 & 2.4 &
93.2 & 94.5 & 8 &
N/A & N/A & N/A & N/A &
0 & 99.9 & 40.1
\\ \hline

c1908 & 97.5 & 52.3 & 4.3 &
91 & 97.8 & 17.7 &
70.8 & 83.9 & 81.9 & 79.9 &
0 & 99.9 & 46.2
\\ \hline

c2670 & 86.3 & 99.9 & 7 &
86.3 & 100 & 7.5 &
52.8 & 66.6 & 66.9 & 56.5 &
0 & 99.9 & 39.8
\\ \hline

c3540 & 88.2 & 95.4 & 18.2 &
82.6 & 98.8 & 27.9 &
44.8 & 40.3 & 41.7 & 42.4 &
0 & 99.9 & 47.9
\\ \hline

c5315 & 93.5 & 98.7 & 4.3 &
91.1 & 98.7 & 12.5 &
49.5 & 54.1 & 50.1 & 56.2 &
0 & 99.9 & 38.3
\\ \hline

c6288 & 97.8 & 36.8 & 3 &
97.6 & 74.2 & 16.5 &
N/A & N/A & N/A & N/A &
0 & 99.9 & 31.6
\\ \hline

c7552 & 97.8 & 69.5 & 1.6 &
97.9 & 81.7 & 3.1 &
56.9 & 48.9 & 53.3 & 48.5 &
0 & 99.9 & 27.8
\\ \hline

\textbf{Average} & 94.3 & 65.3 & 7.1 &
91.9 & 84.5 & 15.1 &
57.0 & 66.1 & 66.4 & 62.9 &
0 & 99.9 & 40.4
\\ \hline

\end{tabular}

%% file: incl/tab-sec-comp2.tex
\begin{tabular}{|*{16}{c|}}
\hline
\multirow{3}{*}{\textbf{Benchmark}}
& \multicolumn{3}{|c|}{\textbf{Original Layout}} 
& \multicolumn{3}{|c|}{\textbf{Pin Swapping \cite{rajendran13_split}}} 
& \multicolumn{3}{|c|}{\textbf{Routing Perturbation \cite{wang17}}} 
& \multicolumn{3}{|c|}{\textbf{Synergistic SM \cite{feng17}}} 
& \multicolumn{3}{|c|}{\textbf{Proposed}} \\
\cline{2-16}
& \textbf{CCR} & \textbf{OER} & \textbf{HD}
& \textbf{CCR} & \textbf{OER} & \textbf{HD}
& \textbf{CCR} & \textbf{OER} & \textbf{HD}
& \textbf{CCR} & \textbf{OER} & \textbf{HD}
& \textbf{CCR} & \textbf{OER} & \textbf{HD}
\\ \hline
c432 & 92.4 & 75.4 & 23.4 &
92.5 & N/A & 39.8 &
78.8 & 99.4 & 46.1 &
N/A & N/A & N/A &
0 & 99.9 & 48.4
 \\ \hline
 
c880 & 100 & 0 & 0 &
85 & N/A & 26 &
47.5 & 99.9 & 18 &
N/A & N/A & N/A &
0 & 99.9 & 43.4
 \\ \hline
 
c1355 & 95.4 & 59.5 & 2.4 &
86 & N/A & 40 &
77.1 & 100 & 26.6 &
N/A & N/A & N/A &
0 & 99.9 & 40.1
\\ \hline

c1908 & 97.5 & 52.3 & 4.3 &
86.2 & N/A & 25 &
83.8 & 100 & 38.8 &
N/A & N/A & N/A &
0 & 99.9 & 46.2
\\ \hline

c2670 & 86.3 & 99.9 & 7 &
N/A & N/A & N/A &
58.3 & 100 & 14 &
33.3 & N/A & 20.5 &
0 & 99.9 & 39.8
\\ \hline

c3540 & 88.2 & 95.4 & 18.2 &
83.5 & N/A & 50 &
77 & 100 & 36.1 &
11.5 & N/A & 35 &
0 & 99.9 & 47.9
\\ \hline

c5315 & 93.5 & 98.7 & 4.3 &
92.5 & N/A & 41 &
74.7 & 100 & 18.1 &
14.9 & N/A & 23.6 &
0 & 99.9 & 38.3
\\ \hline

c6288 & 97.8 & 36.8 & 3 &
N/A & N/A & N/A &
80.9 & 100 & 42.1 &
33.1 & N/A & 40.6 &
0 & 99.9 & 31.6
\\ \hline

c7552 & 97.8 & 69.5 & 1.6 &
91 & N/A & 48 &
73.9 & 100 & 20.3 &
21.3 & N/A & 24.7 &
0 & 99.9 & 27.8
\\ \hline

\textbf{Average} & 94.3 & 65.3 & 7.1 &
88.1 & N/A & 33.4 &
72.4 & 99.9 & 28.9 & 
20.8 & N/A & 28.9 &
0 & 99.9 & 40.4
\\ \hline

\end{tabular}

%% file: incl/tab-sec-comp-magana17.tex
\begin{tabular}{|c|c|c|c|c|}
\hline
\multirow{3}{*}{\textbf{Benchmark}} & \multicolumn{2}{|c|}{\textbf{Routing Blockage~\cite{magana17}}} & \multicolumn{2}{|c|}{\textbf{Proposed Scheme}} \\
\cline{2-5}
 &
 \textbf{$\Delta_+$V67 (\%)} &
 \textbf{$\Delta_+$V78 (\%)} &
 \textbf{$\Delta_+$V67 (\%)} &
 \textbf{$\Delta_+$V78 (\%)} \\
 \hline \hline
\emph{superblue1} &  
23.28 & 65.07 & 36.32 & 49.22 \\ \hline
\emph{superblue5} &  
12.74 & 24.01 & 55.12  & 59.47 \\ \hline
\emph{superblue10} &  
64.85 & 84.09 & 62.09 & 73.12 \\ \hline
\emph{superblue12} &  
16.99  & 35.59 & 79.34 & 70.59 \\ \hline
\emph{superblue18} &  
24.73 & 58.66 & 61.87  & 124.16 \\ \hline
\hline
\textbf{Average} & 
28.52 & 53.48 & 58.95 & 75.31 \\ \hline
\end{tabular}